\documentclass[prb,twocolumn,preprintnumbers,amsmath,amssymb,superscriptaddress]{revtex4}
 
\usepackage{graphicx}
\usepackage{dcolumn}
\usepackage{bm}
\usepackage{nicefrac}

\usepackage{multirow}
\usepackage{longtable}
\usepackage{color}
 
\begin{document} 
 
\title{Bending modes, elastic constants and mechanical stability of graphitic systems} 

\author{G. Savini} 
\affiliation{
Institute for Molecules and Materials, Radboud University Nijmegen, 6525ED Nijmegen, The Netherlands
}  
\affiliation{
Department of Engineering, University of Cambridge, CB3 0FA Cambridge, United Kingdom
}
 
\author{Y. J. Dappe} 
\affiliation{
Institut de Physique et Chimie des Mat{\'e}riaux, CNRS, F-67034 Strasbourg, France
}  
 
\author{S. \"{O}berg} 
\affiliation{
Department of Mathematics, Lule{\aa} University of Technology, S-97187 Lule{\aa}, Sweden
} 
  
\author{J. -C. Charlier}  	
\affiliation{
Institute of Condensed Matter and Nanosciences (IMCN), Universit{\'e} Catholique de Louvain, Place Croix du Sud 1 (NAPS-Boltzmann), B-1348 Louvain-la-Neuve, Belgium
} 
   
\author{M. I. Katsnelson} 
\affiliation{
Institute for Molecules and Materials, Radboud University Nijmegen, 6525ED Nijmegen, The Netherlands
}
 
\author{A. Fasolino} 
\affiliation{
Institute for Molecules and Materials, Radboud University Nijmegen, 6525ED Nijmegen, The Netherlands
} 
   
  
\begin{abstract} 
The thermodynamic and mechanical properties of graphitic systems are strongly dependent on the 
shear elastic constant $C_{44}$. Using state-of-the-art density functional calculations, 
we provide the first complete determination of their elastic constants and exfoliation energies. 
We show that stacking misorientations lead to a severe lowering of $C_{44}$ of at least one order of magnitude. 
The lower exfoliation energy and the lower $C_{44}$ (more bending modes) suggest that flakes 
with random stacking should be easier to exfoliate than the ones with perfect or rhombohedral 
stacking. We also predict ultralow friction behaviour in turbostratic graphitic systems.  
\end{abstract} 
  
  
\maketitle 
\section{Introduction} 
Graphitic systems are used for a wide variety of industrial applications, ranging from lubricant 
and refractory materials to neutron moderators in nuclear fission reactors\cite{Klimenkov-1962} 
and plasma shields in the next generation of fusion reactors\cite{Evans-2006,Scaffidi-Argentina-2000}. 
The recent realization of graphene\cite{Novoselov-2004} (single graphitic layer) 
and the discovery of its unusual electronic properties\cite{Novoselov-2005,Zhang-2005} have raised 
the interest on flake graphitic systems as a route to produce graphene samples of high quality and 
in large scale\cite{Li-2008,Hernandez-2008,Siegel-2010}.  
 
Despite the technological and scientific importance of graphitic systems, the knowledge of their 
elastic properties is unexpectedly poor and new insights are needed. The values of the elastic 
constants describe the mechanical behaviour\cite{Mohiuddin-2009} and are decisive in engineering design to avoid 
material failure.  In layered materials, they are even more important for the thermodynamic properties 
due to a low-lying branch of acoustic vibrations, the bending modes, predicted by Lifshitz\cite{Lifshitz-1952} over 
fifty years ago. Here we show that the shear elastic constant $C_{44}$ affects the mechanism of exfoliation 
that is relevant for the production of graphene.   
 
The most reliable experimental studies of the elastic constants have been obtained by inelastic x-ray 
scattering\cite{Bosak-2007} and ultrasonic, sonic resonance, and static test methods\cite{Blakslee-1979}. 
The sample used in the first study\cite{Bosak-2007} was single-crystalline Kish graphite, characterized by 
an extraordinary high degree of ordering, the closest approximation to the perfect AB stacking graphite (hex-g). 
The second study\cite{Blakslee-1979} was done using highly oriented pyrolitic graphite, the closest 
approximation to turbostratic graphite (turbo-g) where the graphitic layers are randomly oriented around the c-axis.  
 
Except for $C_{44}$ and $C_{13}$, both studies are in agreement within the experimental uncertainties. 
The $C_{13}$ value in turbo-g was determined only by the less accurate static test method and it may be 
affected by errors. Conversely, $C_{44}$ in turbo-g was determined from the sound velocity and its value 
ranges between 0.18-0.35 GPa\cite{Blakslee-1979}, one order of magnitude lower than 5.0±3 GPa found in hex-g\cite{Bosak-2007}.  
 
The discrepancy on $C_{44}$ is attributed to the existence of mobile basal dislocations\cite{Blakslee-1979,Seldin-1970}. 
After neutron irradiation, the elastic constant $C_{44}$ increases by up to an order of magnitude\cite{Seldin-1970}, 
suggesting that interstitial defects\cite{Suarez-Martinez-2007a} could pin the dislocation motions, whence the intrinsic value 
of $C_{44}$ is measured. A principal difficulty with this explanation is that interstitial atoms inevitably 
increase the shear resistance between graphitic layers and therefore they may increase the $C_{44}$ value 
by themselves. 
 
The aim of this study is to investigate from first principles the elastic constants of graphitic systems 
with respect to the stacking misorientations between layers, and to describe the key role of the shear elastic 
constant $C_{44}$ on the bending modes (thermal property) and mechanical stability. We show that stacking 
misorientations greatly affect $C_{44}$ and that graphitic systems with perfect (hex-g) and random (turbo-g) 
stacking should be considered as two distinct materials described by their own elastic and thermodynamic properties.
 
This paper is structured as follow. In Sec.~\ref{method}, we give a brief summary of the theoretical methods and a 
discussion of the LCAO-S$^2$+vdW formalism to include the long-range van der Waals (vdW) interactions. 
The consequences of shear elastic constant $C_{44}$ on the bending modes and mechanical stability are shown 
in Sec.~\ref{bending_modes} and ~\ref{mechanical_stability}, respectively. 
 
The elastic constants in the case of high-symmetric systems (hexagonal, orthorhombic, rhombohedral 
and AA hexagonal stackings) and for graphitic layers randomly oriented around the c-axis (turbostratic stacking) 
are presented in Sec.~\ref{EC_high-symmetric} and ~\ref{EC_turbostratic}, respectively. 
Finally, we summarize and comment on our results (Sec.~\ref{conclusion}).

\section{Method\label{method}}
 
All the calculations are performed using density functional theory, within the local density approximation 
scheme (LDA), norm-conserving pseudopotentials\cite{Troullier-1991} and plane waves with cut-off energy of 150 Ry 
({\sc abinit} package)\cite{Gonze-2002}. The k-point mesh was chosen so that the average density corresponds approximately to 
a 32x32x16 mesh for hex-g. Energies were converged within 0.05 meV/atom, and elastic constants 
within 0.5\%. For large supercells (turbo-g with more than 50 atoms) we have used localized basis-set 
composed of Gaussian orbitals ({\sc aimpro} code)\cite{Briddon-2000}. The elastic constants calculated by the two 
LDA codes are in agreement within 3\% or better. 
 
The choice of LDA is not fortuitous\cite{Perdew-1981} and it was dictated by test calculations using the generalized-gradient 
approximation (GGA) within the Perdew-Burke-Ernzerhof scheme\cite{Perdew-1996}. According to GGA, the distance between graphitic 
layers is far too large (4.2 {\AA}), resulting in a negligible interlayer binding energy and almost vanishing out-of-plane elastic constants ($C_{44}$, $C_{33}$). For these reasons we have dismissed the use of GGA from 
this study\cite{Tournus-2005,Charlier-1994a,comment}. 
 
Even though LDA yields accurate equilibrium distances, due to energetical error compensations, the long-range 
van der Waals (vdW) and more generally the weak contribution to the out-of-plane interactions is not well 
described\cite{Mounet-2005,Charlier-1994b}. In order to check the importance of these effects on $C_{44}$ and $C_{33}$, 
we have used the LCAO-S$^2$+vdW formalism to include these specific interactions within LDA\cite{Dappe-2006} 
(using {\sc fireball} code)\cite{Lewis-2001}. 
 
This formalism takes into account two major contributions. The first one, which we call weak chemical interaction, 
is a repulsive energy originating from the overlaps of electronic densities between the weakly interacting subsystems. 
Even though the overlaps are rather small, this energy is not negligible. This contribution is evaluated 
proceeding to a second order expansion of the electronic wavefunctions with respect to the overlaps.  
 
The second contribution, which is the vdW itself, originating from charge fluctuations, can be seen as the 
interaction between electronic dipoles. In the frame of the dipolar approximation, we use a second order 
perturbation theory to describe this contribution. This method has originally been tested with success on 
standard graphene-graphene interaction, and more recently on a wide range of graphitic materials\cite{Dappe-2009}. 
In its current stage, the analysis of the internal stresses is not implemented yet, thus the elastic constants that 
change the in-plane bond lengths ($C_{11}$, $C_{12}$, $C_{13}$) are overestimated by up to 15\%. However, 
for $C_{44}$ and $C_{33}$ that describe the weak interaction between layers, the internal stresses are negligible, 
and the calculated values are expected to be very accurate. 
 
The elastic constants are determined using two different approaches. The first one uses the response-function, 
implemented in the {\sc abinit} code, to calculate the second derivative of the total energy with respect to the strains. 
 
The second approach uses the elastic energy density\cite{Born-1954}. For each elastic constant we have applied 21 strain 
components $\varepsilon_{ij}$ to the equilibrium crystal structures ($\varepsilon_{ij}$ were typically ranging 
between ±0.01 with increments 0.001). The elastic constants are found by fitting the calculated energies to a 
polynomial function in the strains. Both approaches yield results in agreement within 0.2\%.
 
\section{Results and discussion} 
\subsection{Bending modes\label{bending_modes}}
The bending modes are atomic vibrations that can be excited even at low temperature and strongly influence the 
thermal properties of layered materials. In 1952, Lifshitz\cite{Lifshitz-1952} obtained the following dispersion law for the 
out-of-plane acoustic mode $\omega$: 
\begin{equation} 
\rho\times\omega^{2}\left(q\right)=C_{44}\left(q_{x}^{2}+q_{y}^{2}\right)+C_{33}q_{z}^{2}+\kappa\left(q_{x}^{2}+q_{y}^{2}\right)^{2}/c
\end{equation} 
where $\rho$ is the density, $c$ is the interlayer distance, $q_{x,y,z}$ are the wave vectors and $\kappa$ describes 
the intralayer forces characterizing the bending rigidity (1.1 eV)\cite{Fasolino-2007}. 
 
\begin{figure}[t]
\includegraphics[width=8.5cm]{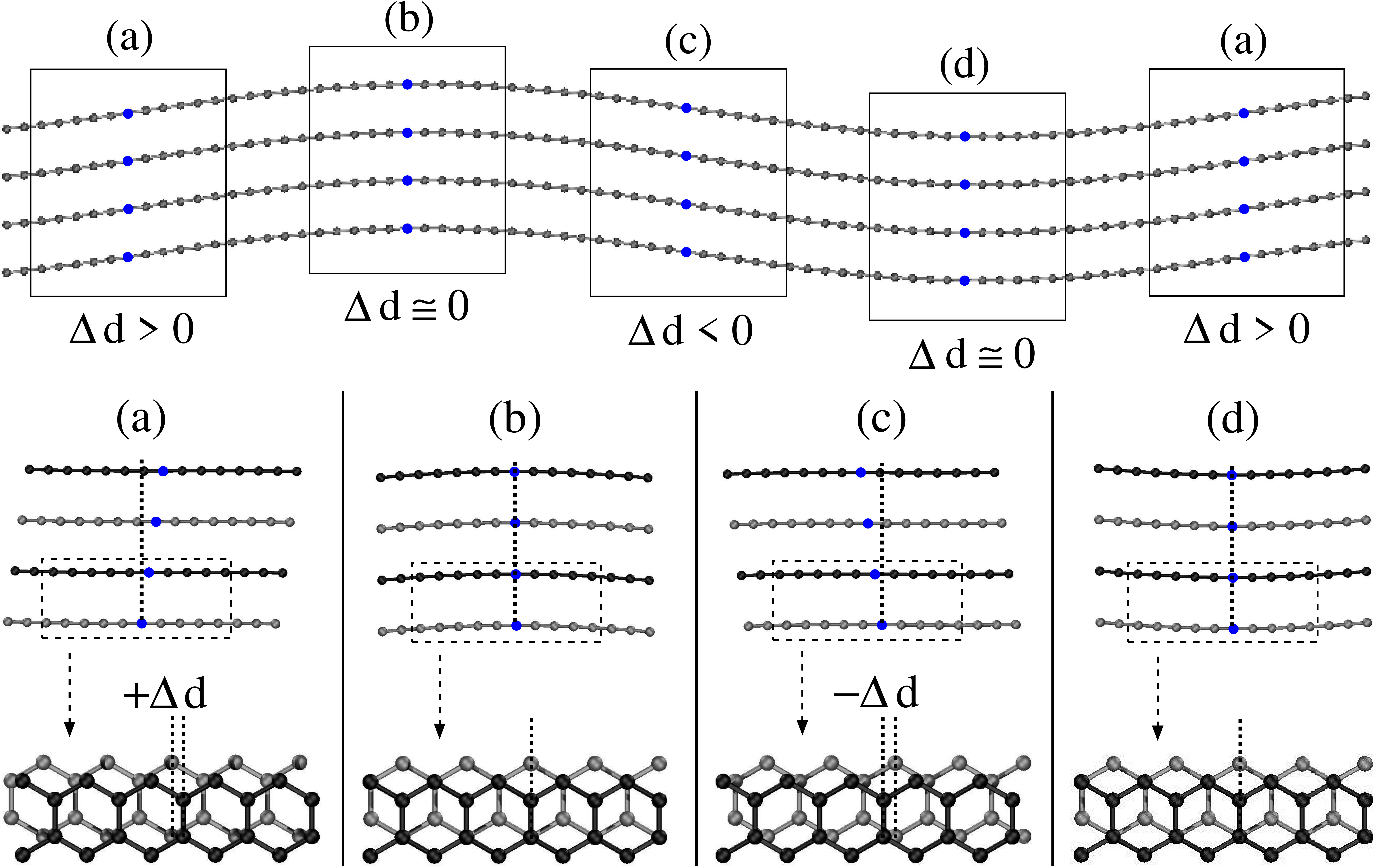}
\caption{\label{Figure1} (Color online) 
Transversal acoustic (bending) mode. The bending changes the local stacking between graphitic layers. 
The boxes (a-d) show regions with different slopes and stackings. The shear $\triangle d$ gives the deviations 
from AB phase (perfect stacking). } 
\end{figure} 
 
The small value of $C_{44}$ (characteristic of graphitic systems, see Table~\ref{Table:elastic_constants}) 
leads to a predominant contribution of the transversal bending modes ($q_{z}=0)$) in 
the phonon dispersion curves. These modes are sinusoidal displacements that propagate along the planes and change the 
local stacking between layers (see Fig.~\ref{Figure1}). For nearly flat planes the shear stacking $\triangle d$ is 
almost zero, whereas, for positive (or negative) slope $\triangle d$ becomes positive (or negative). 
Using trigonometric considerations, it can be shown that the maximum value of $\triangle d$ (see boxes in Fig.~\ref{Figure1}) 
is given by: 
\begin{equation} 
\triangle d=\frac{\pi\widehat{a}c}{\lambda}\times\frac{1}{\sqrt{1+\left(\frac{2\pi\cdot\widehat{a}}{\lambda}\right)^{2}}}
\end{equation} 
where $\widehat{a}$ is the amplitude, $\lambda$ is the wavelength and $c$ is the interlayer distance. 
 
The crystal resistance to the stacking shear\cite{Born-1954} is proportional to $E\propto C_{44}\times\triangle d^{2}$, 
and by lowering $C_{44}$ more bending modes can be excited at lower temperature. In the limit of $C_{44}=0$, 
graphitic systems approach the graphene behaviour, where indeed bending modes (or ripples) are 
always present\cite{Fasolino-2007}.
 
\subsection{Mechanical stability\label{mechanical_stability}} 
 
By imposing the elastic strain energy as positively definite\cite{Born-1954}, the stability conditions are given by:  
\begin{equation} 
2C_{13}^{2}<C_{33}\left(C_{11}+C_{12}\right)\:\:\:\:\: C_{11},C_{12},C_{33},C_{44}>0 
\end{equation} 
Note that $C_{13}$ does not affect the stability: a positive (or negative) value means that under in-plane compression 
the out-of-plane distance tends to expand (or contract). 
 
The elastic constants $C_{11}$, $C_{12}$ describe in-plane deformations and they possess the highest values due to the 
strong sp$^2$ bonding interactions within the graphene planes. The elastic constant $C_{33}$ describes out-of-plane 
compression or expansion and it has always a positive value for perfect and stacking misorientations (see 
Table~\ref{Table:elastic_constants}). 
 
The elastic constant $C_{44}$ corresponds to a shear between graphene layers. Due to the weak interaction between planes, 
the $C_{44}$ value is the lowest and can be positive or negative depending on the stacking misorientations 
(see Table~\ref{Table:elastic_constants}). The latter elastic constant is the only one that can break the 
mechanical stability condition (\emph{i.e.} $C_{44}<0$).
 
\subsection{Elastic constants in high-symmetric \\ graphitic systems\label{EC_high-symmetric}}
 
\begin{figure}[t]
\includegraphics[width=8.5cm]{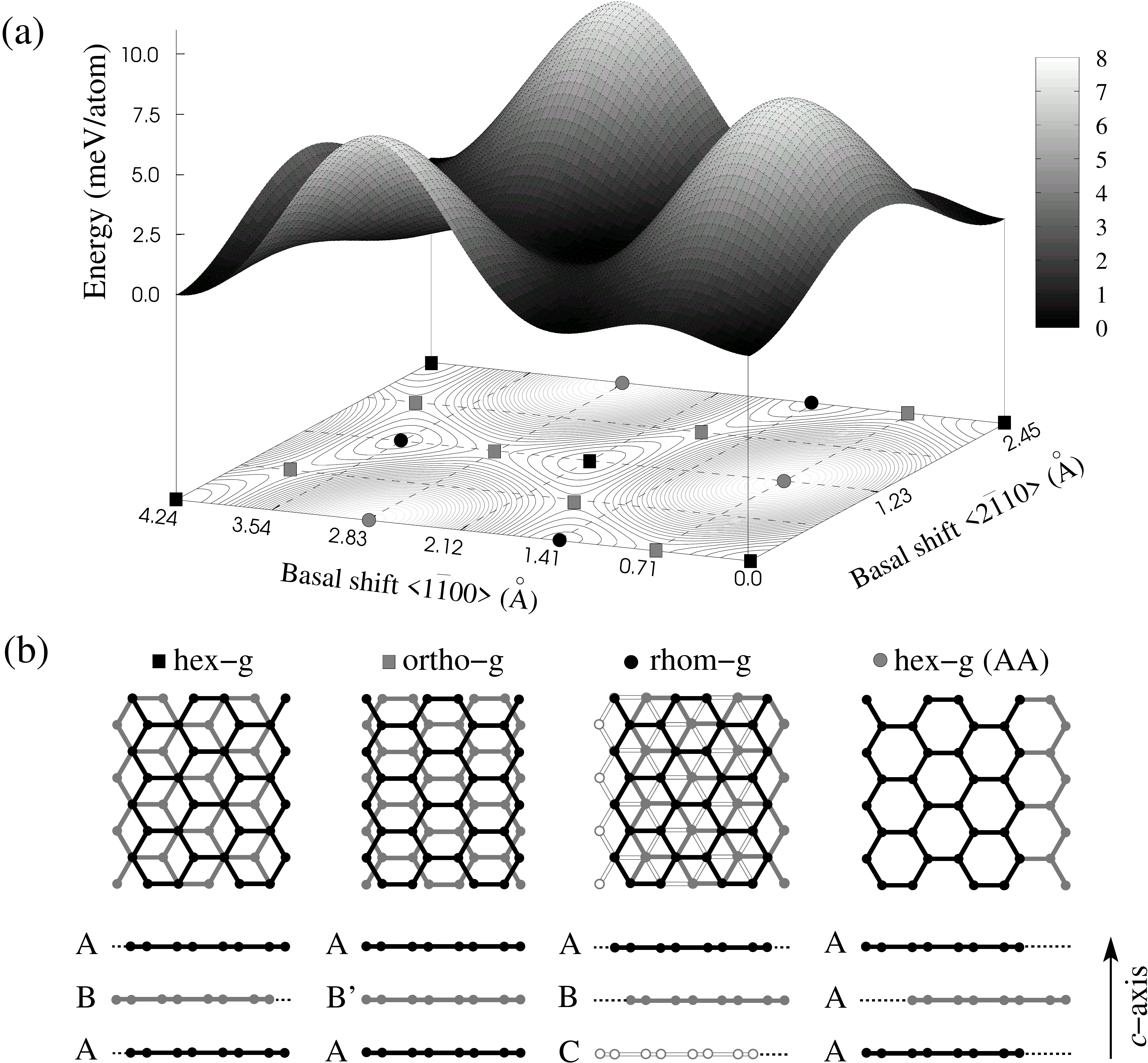}
\caption{\label{Figure2}
(a) Stacking-fault energy surface (also called corrugation energy surface). The square and circle symbols indicate 
the stationary points corresponding to the following high-symmetric structures; 
(b) The hexagonal, orthorhombic, rhombohedral and AA hexagonal stackings viewed perpendicular (above), parallel (below) 
to the c-axis. The energies of these structures with respect to hex-g are 1.66 meV/atom for ortho-g, 0.10 meV/atom for 
the rhom-g and 9.29 meV/atom for AA hex-g (see text). 
}  
\end{figure}

By imposing a translation vector between graphitic layers we found four high-symmetric stackings that correspond 
to stationary points on the stacking-fault energy surface\cite{Kaxiras-1993}. To calculate the stacking-fault energy surface 
(for graphitic system is often called corrugation energy surface)\cite{Kolmogorov-2005,Kolmogorov-2000,Kolmogorov-2004}, 
we have used 16-atoms unit cell model of eight layers in AB sequence in which the stacking at the unit cell 
boundary is changed by imposing a shear displacement. This represents an intrinsic stacking fault between layers 
at the unit cell boundaries whereas the others remains stacked in the AB sequence. 
The multiple layers repeat along the c-axis (8 layers) makes negligible 
self-interaction of the intrinsic fault\cite{Kaxiras-1993}. The stationary points, indicated with square and 
circle symbols in Fig.~\ref{Figure2}a, correspond to the four high symmetric structures (hexagonal, rhombohedral, 
orthorhombic and AA hexagonal graphite, see Fig.~\ref{Figure2}b). The in-plane lattice parameter was 2.45 {\AA} 
very close to the experimental value 2.463{\AA}\cite{Bosak-2007}, with no significant changes among all the graphitic structures.  
 
The origin in Fig.~\ref{Figure2}a corresponds to hex-g (black square symbol), the global minimum of the energy 
surface with an interlayer separation of 3.34 {\AA} (experimental value 3.356 {\AA})\cite{Bosak-2007}. With the exception 
of the $C_{33}$ value (29 GPa), the calculated elastic constants (given in Table~\ref{Table:elastic_constants}) are 
in agreement within the experimental uncertainties found in hex-g\cite{Bosak-2007}. Using the LCAO-S$^2$+vdW formalism, 
the out-of-plane elastic constant $C_{33}$ becomes 42 GPa, very close to the experimental value 38.7$\pm$7 GPa\cite{Bosak-2007}.  
 
The local minimum at $\nicefrac{1}{3}$$\langle$$1\overline{1}00$$\rangle$ (black circle symbol) represents the 
rhombohedral stacking (rhombo-g). This structure possesses the same interlayer separation and elastic constants of hex-g 
(the differences are beyond the accuracy of the calculations) with formation energy of 0.10 meV/atom. This very small 
energy explains why the rhombohedral phase is usually 5-15\% intermixed with the perfect hexagonal one in natural 
graphitic flakes.  
 
The saddle point at $\nicefrac{1}{6}$$\langle$$1\overline{1}00$$\rangle$ (grey square symbol) represents the orthorhombic 
stacking (ortho-g). The interlayer separation of the primitive unit cell is 3.37 {\AA} with formation energy of 1.66 meV/atom. 
The latter energy represents the lowest barrier that has to be overcome during the shearing process from an ideal 
configuration to another equivalent one. The $C_{44}$ values are found to range between -2.7 GPa, 
for shearing along $\langle$$1\overline{1}00$$\rangle$, and 7.7 GPa, for shearing along the $\langle$$2\overline{1}10$$\rangle$ axis. 
No significant changes are found for the other elastic constants (see Table~\ref{Table:elastic_constants}). 
Although unstable (due to the negative $C_{44}$), this structure acts as an intermediate phase during the transformation 
from graphite to diamond\cite{Scandolo-1995}. 
 
The global maximum at $\nicefrac{2}{3}$$\langle$$1\overline{1}00$$\rangle$ (grey circle symbol) corresponds to AA 
hexagonal stacking (AA hex-g). This phase has the largest interlayer separation (3.60 {\AA}) and the highest formation 
energy (9.29 meV/atom). Its elastic constants are smaller than in hex-g, and $C_{44}$ in particular, becomes 
negative (-3.8 GPa) breaking the stability conditions. Even though this structure is highly unstable, a recent study has 
suggested that screw dislocations locally encourage this stacking\cite{Suarez-Martinez-2007b}.  
 
In the following section we describe the elastic constants for graphitic layers randomly oriented around the c-axis. 
 
\subsection{Elastic constants in turbostratic stacking\label{EC_turbostratic}}
 
The modelling of turbostratic staking is challenging since the incommensurate nature of these stackings must combine with 
the finite-size constraint required by calculations. To overcome this difficulty we used the method proposed by Kolmogorov 
and Crespi\cite{Kolmogorov-2005} in which a layer with supercell basis vector ${(n,m)}$ becomes commensurate with a second layer for 
a specific rotation angle of: 
\begin{equation}
\vartheta=\cos^{-1}\left[\frac{2 n^{2}+2 nm -m^{2}}{2\left(n^{2}+nm+m^{2}\right)}\right]
\qquad \textrm{with}\;\; n>m
\end{equation}
 
\begin{figure}[t]
\includegraphics[width=7.5cm]{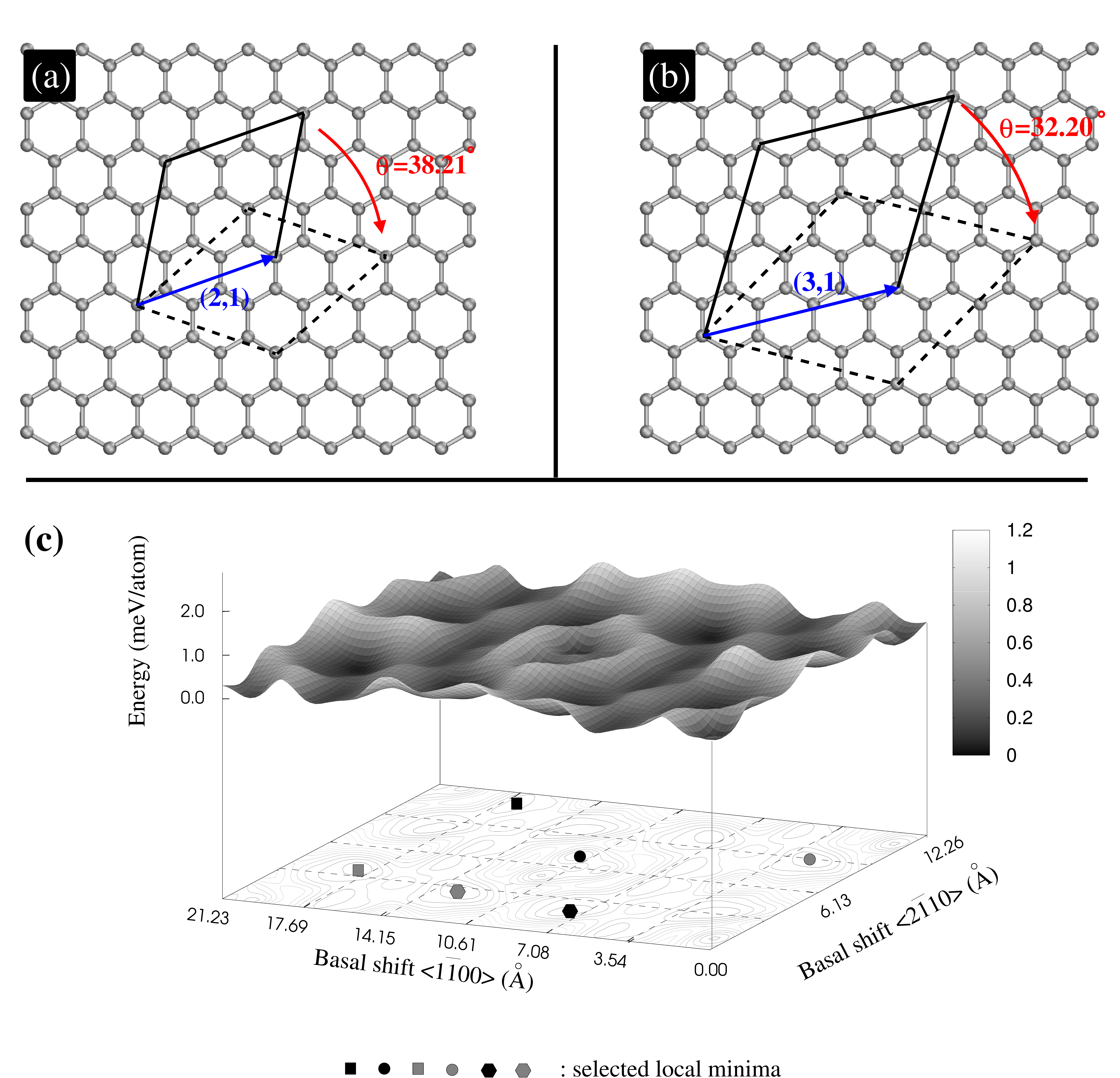}
\caption{\label{Figure3}(Color online) 
(a,b) Representation of accidental commensuration. A supercell of vector $(n,m)$ (blue colour) becomes commensurate 
when rotated by an angle $\theta$ with respect to the starting supercell (as we increase $n,m$ the supercell surface 
and number of atoms rapidly increase). (c) Stacking fault energy surface (or corrugation energy surface) of a 
graphene bilayer with supercell vector (2,1). The respective elastic constants are calculated on the local minima 
of the energy surface. Notice that the energy surface becomes much flatter leading to a reduction of $C_{44}$. } 
\end{figure}
 
\begin{figure}[t]
\includegraphics[width=8.5cm]{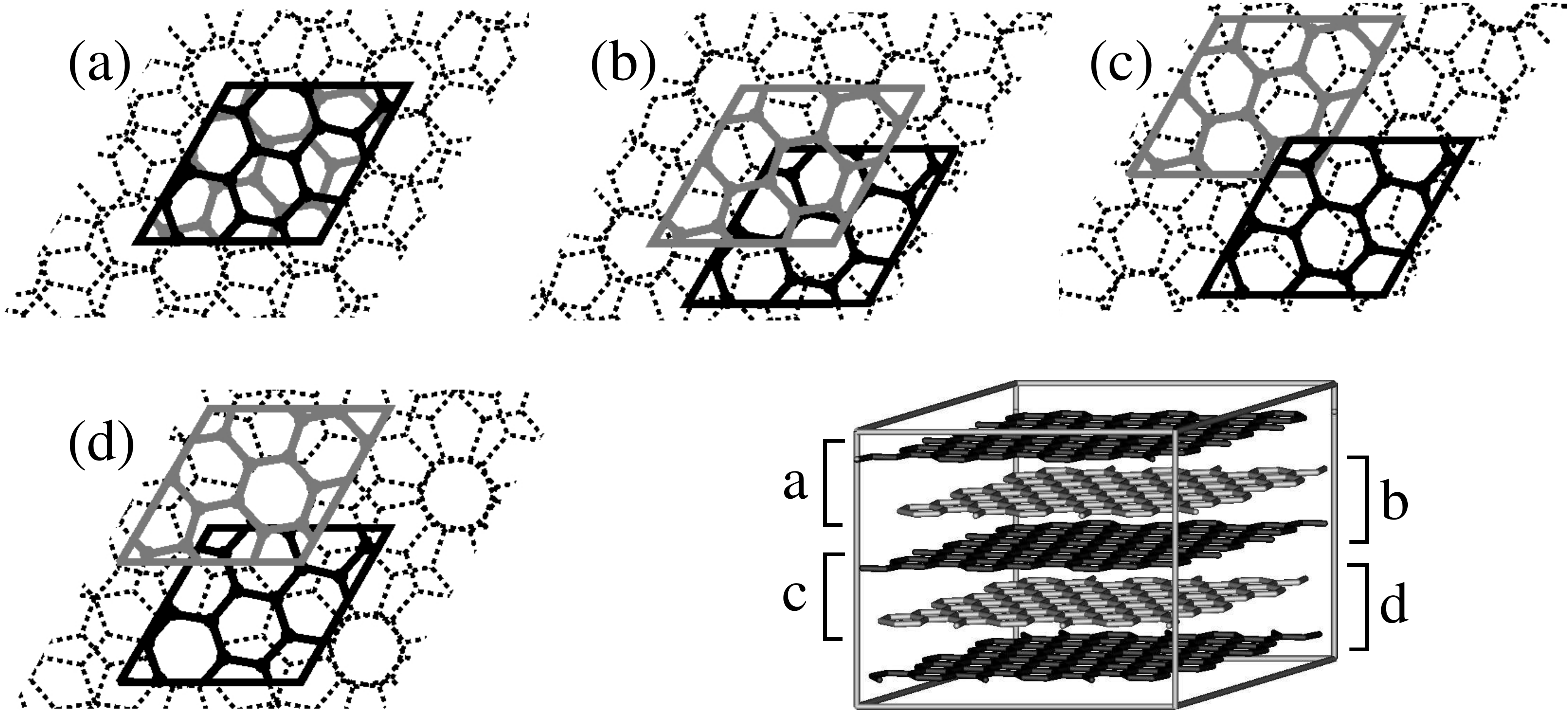}
\caption{\label{Figure4} 
An example of 5-layers turbostratic stacking (supercell basis vector 2,1). Each layer is stacked along 
the c-axis, rotated with respect to each other with an angle of 38.21$^{\circ}$ and randomly translated along 
the basal plane.
}
\end{figure}

Figure~\ref{Figure3}a,b shows the case of the two smaller supercell with basis vector $(2,1)$ and $(3,1)$ 
(corresponding to 14, 26 atoms/layer and rotation angles of 38.21$^{\circ}$, 32.20$^{\circ}$, respectively). 
Figure~\ref{Figure3}c shows the stacking energy surface for a bilayer supercell of basis vector $(2,1)$. To better 
describe the complete misorientation of turbostratic stacking we have increased the number of layers along the c-axis. 
Each layer was rotated with respect to each other and randomly translated along the basal plane (see Fig.~\ref{Figure4}). 
The rotation angles used are 15 values ranging from 6.01$^{\circ}$ to 53.99$^{\circ}$. The smallest supercell 
contains 28 atoms with 2 layers rotated with respect to each other by 38.21$^{\circ}$, whereas the largest one 
contains 456 atoms with 12 layers and rotation angle of 46.83$^{\circ}$. 
 
For each model we carried out extensive structural optimizations starting from different translation vectors along 
the basal plane. The corrugation energy is about one order of magnitude lower than commensurate structure 
(see Fig.~\ref{Figure2}a) with a maximum value of 0.9 meV/atom (see Fig.~\ref{Figure3}c). 
As we increased the size of the supercell over the basal plane we found that the average corrugation energy tends 
to decrease up to 20\% for the largest size (basis vector 8,3 with 194 atoms/layer). Extrapolating these results 
in the ideal case of infinite layers, we suggest corrugation energy virtually flat with layers mutual independent. 
 
For all the supercell studied the in-plane lattice parameters remain almost equal to the value found in hex-g, 
whereas the interlayer distances were on average slightly larger, 3.42$\pm$0.01 {\AA}, with formations energies 
of 3.03$\pm$0.6 meV/atom. 
 
The interlayer binding energy between graphitic layers (\emph{i.e.} exfoliation energy) was 21$\pm$1 meV/atom 
(70$\pm$4 meV/atom with vdW), a value slightly lower than the 24 meV/atom found in hex-g (80 meV/atom with vdW). 
Note that LDA values yield to a binding energy within a factor of 2-3 with respect to LCAO-S$^2$+vdW formalism and 
experiment values (43 meV/atom found in heat-of-wetting experiment\cite{Girifalco-1956}, 35$\pm$10 meV/atom found 
by analyzing TEM images of twisted collapsed nanotubes\cite{Benedict-1998}, and 52$\pm$5 meV/atom by studying 
thermal desorption of polyaromatic hydrocarbons\cite{Zacharia-2004}). In Table~\ref{Table:elastic_constants} 
we report the calculated values of the elastic constants. 
With the exception of $C_{44}$, these values hardly change among all the studied turbostratic stackings with no 
clear dependence on the rotation angles and number of layers. The elastic constants $C_{11}$, $C_{12}$ slightly 
decrease by about 3\% with respect to hex-g, remaining within the experimental uncertainties found in turbostratic 
samples\cite{Blakslee-1979}. As previously found in hex-g, LDA calculations underestimate the $C_{33}$ value (27$\pm$2 GPa) 
with respect to the vdW correction (36$\pm$1 GPa) and the experimental value of 36.5$\pm$1GPa \cite{Blakslee-1979}.  
 
Conversely, we found $C_{13}$=-2.7$\pm$0.5 GPa in disagreement with the experimental value 15$\pm$5 GPa\cite{Blakslee-1979}. 
The latter value was only indirectly obtained as a function of the other elastic constants by the less accurate static 
test method. This method requires larger strains than ultrasonic experiments and non-linear behaviour of stress-strain 
curve may affect the measured value. Furthermore, the linear bulk modulus $B_{a}$ calculated from these elastic 
constants is far too large (2080 GPa), almost double than the one found in diamond (1326 GPa)\cite{Grimsditch-1975}. 
 
\begin{figure}[t]
\includegraphics[width=8.3cm]{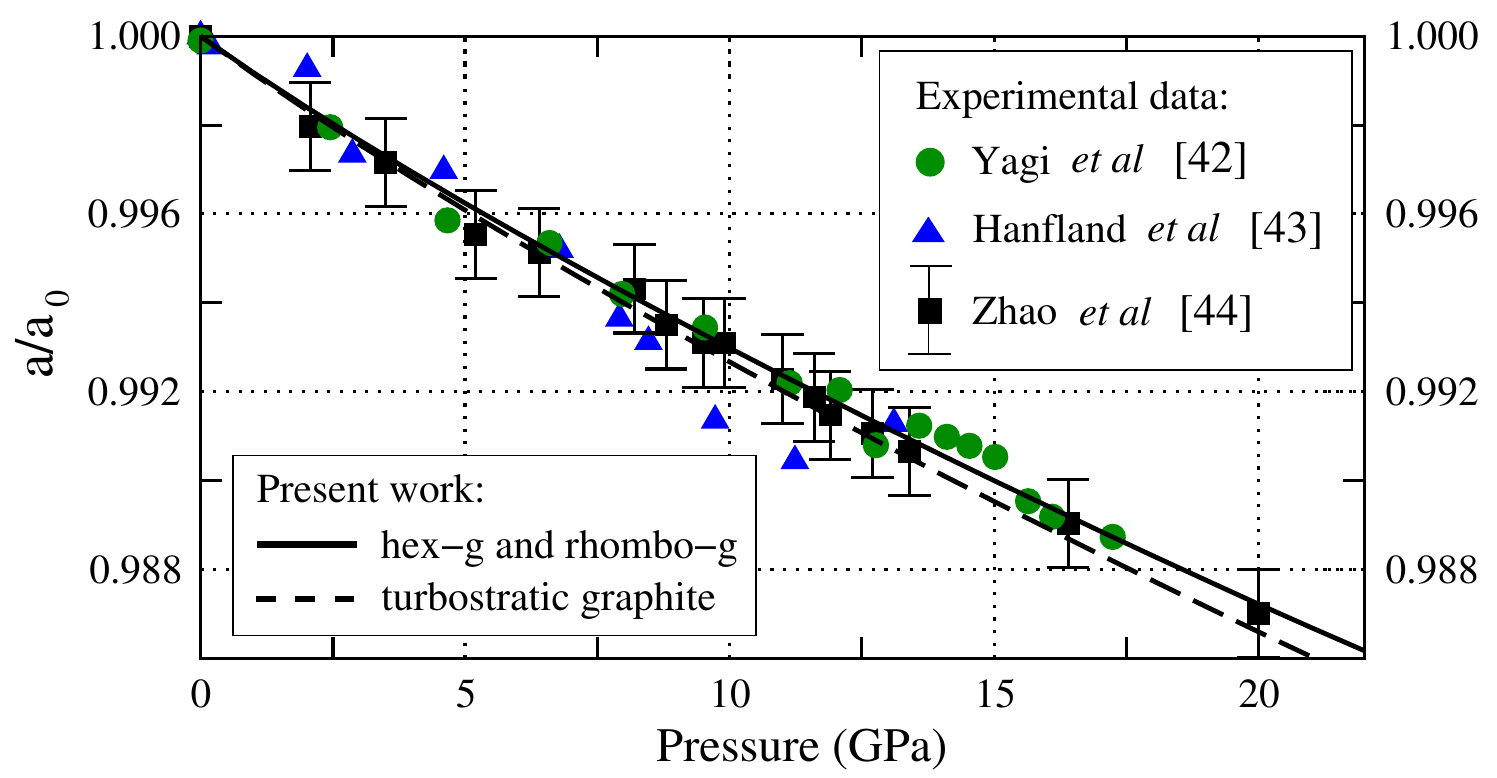}
\caption{\label{Figure5} (Color online) 
In-plane lattice parameters {\em vs.} pressure. The solid line represents the results found for hex-g and rhombo-g 
(perfect and rhombohedral stackings). The dashed line shows the results found here for turbostratic graphite. 
For comparison, the experimental results are also plotted.
} 
\end{figure}
 
The linear bulk modulus $B_{a}$ describes the variation of the lattice parameter $a$ as a function of the hydrostatic 
pressure\cite{Schreiber-1973} and it is given by: 
\begin{equation}
B_{a}=\frac{C_{33}\left(C_{11}+C_{12}\right)-2C_{13}^{2}}{C_{33}-C_{13}} 
\label{eq:linear_bulk_modulus}
\end{equation}
This modulus is strongly weighted by $C_{13}$. For example, if we use the measured value found in hex-g 0$\pm$3 GPa 
(instead of 15$\pm$5 GPa), $B_{a}$ becomes 1240 GPa (instead of 2080 GPa). 
 
Several x-ray studies have measured the linear bulk modulus $B_{a}$. In these experiments powder samples were 
prepared by grinding different types of graphitic materials, ranging from well crystallized to poorly crystallized 
grains\cite{Yagi-1992,Hanfland-1989,Zhao-1989}. The good agreement between the experiments (see Fig.~\ref{Figure5}) 
indicates that the linear bulk modulus $B_{a}$ does not strongly depend on the stacking order with a measured value 
of 1250$\pm$70 GPa\cite{Hanfland-1989}. 
By including the latter modulus in the Eq.~\ref{eq:linear_bulk_modulus}, the elastic constant $C_{13}$ becomes 0.3 GPa. 
Therefore, we conclude that the $C_{13}$ value does not significantly change between turbo-g and hex-g and we 
propose that the same value 0$\pm$3 GPa should be appropriate also for turbostratic stacking. 
 
\begin{figure}[t]
\includegraphics[width=7.6cm]{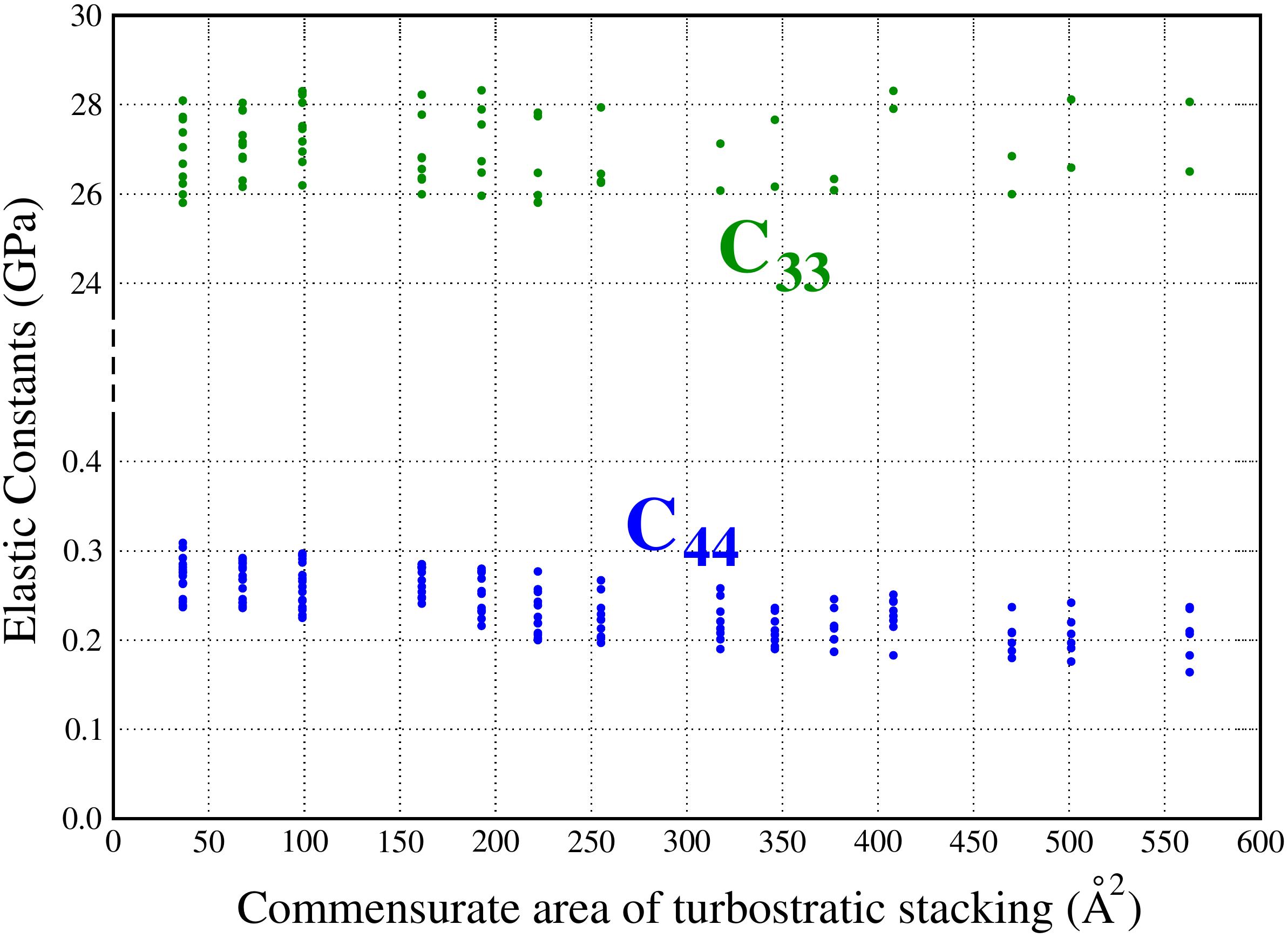}
\caption{\label{Figure6}(Color online) 
The elastic constants $C_{44}$ and $C_{33}$ as a function of the commensurate area over the basal plane in 
turbostratic stacking. The $C_{44}$ value tends to decrease with respect to the supercell size, whereas $C_{33}$ 
is not quantitatively affected. } 
\end{figure}

The $C_{44}$ represents the second derivative of the total energy as a function of shear displacement over the 
basal plane. As the corrugation energy of perfect AB stacking is much higher than turbostratic we are expecting 
a corresponding lower value of the shear elastic constant. We have found that $C_{44}$ tends to decrease as a 
function of the supercell area over the basal plane (see Fig.~\ref{Figure6}) and is independent on the relative 
rotation angles and number of layers along the c-axis (the respective average corrugation energies and therefore 
the $C_{44}$ values are also independent on the relative rotation angles and number of layers). 
 
For commensurate structures of finite area ranging between 36-563 $\textrm{\AA}^{2}$ the respective $C_{44}$ values 
are within 0.16-0.31 GPa, one order of magnitude lower than hex-g (4.5 GPa) and close to the experimental measures 
ranging from 0.18 to 0.35 GPa\cite{Blakslee-1979}. 
The vdW correction does not significantly affect the $C_{44}$ values with respect to LDA 
(see Table~\ref{Table:elastic_constants}), suggesting that the variations in energy under interlayer shears are 
nearly identical between these two approximations. As we increase the sizes of the commensurate structures towards 
the ideal case of infinite layers (incommensurate) the $C_{44}$ values tend to zero with corrugation energy virtually 
flat. These results indicate that turbostratic stacking possesses the lowest friction among all the graphitic 
materials bearing great potential applications in nano-mechanical systems\cite{Dienwiebel-2004,Cumings-2000}. 
 
\begin{table*}[tp]
\begin{centering}
\caption{\label{Table:elastic_constants}
Elastic constants in unit of GPa for the different graphitic systems. The values between brackets are 
calculated using the LCAO-S$^2$+vdW formalism. These results show that the $C_{13}$ values do not significantly 
change between turbo-g and hex-g and we propose that the same value 0$\pm$3 GPa should be appropriate also for 
turbostratic stacking. The shear elastic constants $C_{44}$ found in turbostratic stacking correspond to 
commensurate structures of area ranging between 36-563 $\textrm{\AA}^{2}$. 
} 
\begin{tabular}{|ll|lrlllrcc|crclcrclc|rrll|lrrlc|crl|}
\hline
\multicolumn{1}{|l}{} &  & \multicolumn{8}{c|}{hex-g (AB)} & \multicolumn{9}{c|}{turbo-g} & \multicolumn{1}{c}{} & \multicolumn{2}{c}{rhombo-g} &  & \multicolumn{5}{c|}{ortho-g} & \multicolumn{3}{c|}{hex-g (AA)}\tabularnewline
\cline{3-3} \cline{4-6} \cline{7-9} \cline{10-10} \cline{11-11} \cline{12-14} \cline{15-18} \cline{19-19} \cline{20-20} \cline{21-22} \cline{23-23} \cline{24-28} \cline{29-31} 
\multicolumn{1}{|l}{} & \multicolumn{1}{l|}{} & \multicolumn{1}{l}{} & \multicolumn{3}{l}{Experiment[\onlinecite{Bosak-2007}]} & \multicolumn{3}{c}{Theory} &  &  & \multicolumn{3}{c}{Experiment[\onlinecite{Blakslee-1979}]} & \multicolumn{4}{r}{Theory} &  & \multicolumn{1}{c}{} & \multicolumn{2}{c}{Theory} &  & \multicolumn{5}{c|}{Theory} & \multicolumn{3}{c|}{Theory}\tabularnewline
\hline 
\multicolumn{2}{|l|}{$C_{11}$} &  & $1109$ & $\pm$ & $16$ &  & $1109$ &  &  &  & $1060$ & $\pm$ & $20$ &  & $1080$ & $\pm$ & $3$ &  &  & $1107$ &  &  &  & $1095$ &  &  &  & \multicolumn{2}{r}{$1028$} & \tabularnewline
\multicolumn{2}{|l|}{$C_{12}$} &  & $139$ & $\pm$ & $36$ &  & $175$ &  &  &  & $180$ & $\pm$ & $20$ &  & $171$ & $\pm$ & $4$ &  &  & $175$ &  &  &  & $173$ &  &  &  &  & $162$ & \tabularnewline
\multicolumn{2}{|l|}{$C_{33}$} &  & $38.7$ & $\pm$ & $7$ &  & $29$ & \multicolumn{2}{c|}{$\left(42\right)$} &  & $36.5$ & $\pm$ & $1$ &  & $27$ & $\pm$ & \multicolumn{2}{l|}{$2$ $\left(36\pm1\right)$} &  & $29$ & \multicolumn{2}{l|}{$\left(42\right)$} &  & $26$ & \multicolumn{3}{l|}{$\left(38\right)$} &  & $21$ & $\left(30\right)$\tabularnewline
\multicolumn{2}{|l|}{$C_{13}$} &  & $0$ & $\pm$ & $3$ &  & $-2.5$ &  &  &  & $15$ & $\pm$ & $5$  &  & $-2.7$ & $\pm$ & $1$ &  &  & $-2.5$ &  &  &  & $-2.6$ &  &  &  &  & $-3.0$ & \tabularnewline
\multicolumn{2}{|l|}{$C_{44}$} &  & $5.0$ & $\pm$ & $3.0$ &  & $4.5$ & \multicolumn{2}{c|}{$\left(4.8\right)$} &  & $0.18$ & $/$ & $0.35$ &  & $0.16$ & $/$ & \multicolumn{2}{l|}{$0.31$ $\left(0.19 / 0.34\right)$} &  & $4.4$ & \multicolumn{2}{l|}{$\left(4.8\right)$} & \multicolumn{5}{l|}{$-2.7/7.7$ $(-2.9/7.3)$} &  & $-3.8$ & $($$-3.8)$\tabularnewline
\hline
\end{tabular}
\par\end{centering}
\end{table*}
 
\section{Conclusions\label{conclusion}}
 
To summarize, we have discussed the importance of $C_{44}$ as the main parameter that restrains the bending 
modes and controls the mechanical stability of layered materials. Using advanced \emph{ab-initio} method, which 
includes vdW interactions, we have provided the first complete description of the elastic constants in graphitic systems. 
The lower exfoliation energy (3-8 meV/atom) and the lower $C_{44}$ (at least one order of magnitude) found in 
turbostratic stacking suggest that the exfoliation mechanism, relevant for the production of graphene, should 
be easier for graphite flakes with random stacking. 
Our results indicate that turbostratic graphitic systems possess the lowest friction among all the 
graphitic stackings. It would be interesting to check these predictions experimentally.

\begin{acknowledgments}
The authors thank the HPC-EUROPA project for financial and computer supports. 
Computational resources have also been provided by the Universit{\'e} Catholique de Louvain on the LEMAITRE and GREEN computers 
of the CISM. J.-C.C. acknowledges financial support from the F.R.S.-FNRS of Belgium; G.S. from JSPS and Grant-in-Aid 
for Scientific Research.
\end{acknowledgments}


\begin{thebibliography}{99}
\bibitem{comment} It is clear that GGA (leading to a distance of 4.2 {\AA} and energy of 2-3 meV/atom) seriously underestimates the interlayer binding energy and is in strong disagreement for the geometry. Thus, even if LDA is unable to describe the dispersion interactions, it is a better choice than the GGA as far as the geometry and energy differences are concerned.
\end{thebibliography}
\end{document}